\documentclass[osajnl,twocolumn,showpacs,superscriptaddress,10pt]{revtex4-1} 
\usepackage{amsmath,amssymb,graphicx}
\providehyphenmins{english}{{3}{3}}
\usepackage[english]{babel}

\begin{document}

\title{Frequency comb up- and down-conversion in a synchronously-driven $\chi^{(2)}$ optical microresonator}

\author{Simon J. Herr}
\affiliation{Laboratory for Optical Systems, Department of Microsystems Engineering - IMTEK, University of Freiburg, Georges-K\"{o}hler-Allee 102, D-79110 Freiburg, Germany}

\author{Victor Brasch}
\affiliation{Swiss Center for Electronics and Microtechnology (CSEM), Time and Frequency, Rue de l'Observatoire 58, CH-2000 Neuch\^{a}tel, Switzerland}

\author{Jan Szabados}
\affiliation{Laboratory for Optical Systems, Department of Microsystems Engineering - IMTEK, University of Freiburg, Georges-K\"{o}hler-Allee 102, D-79110 Freiburg, Germany}

\author{Ewelina Obrzud}
\affiliation{Swiss Center for Electronics and Microtechnology (CSEM), Time and Frequency, Rue de l'Observatoire 58, CH-2000 Neuch\^{a}tel, Switzerland}
\affiliation{Geneva Observatory, University of Geneva, Chemin des Maillettes 51, CH-1290 Versoix, Switzerland}

\author{Yuechen Jia}
\affiliation{Laboratory for Optical Systems, Department of Microsystems Engineering - IMTEK, University of Freiburg, Georges-K\"{o}hler-Allee 102, D-79110 Freiburg, Germany}

\author{Steve Lecomte}
\affiliation{Swiss Center for Electronics and Microtechnology (CSEM), Time and Frequency, Rue de l'Observatoire 58, CH-2000 Neuch\^{a}tel, Switzerland}

\author{Karsten Buse}
\affiliation{Laboratory for Optical Systems, Department of Microsystems Engineering - IMTEK, University of Freiburg, Georges-K\"{o}hler-Allee 102, D-79110 Freiburg, Germany}
\affiliation{Fraunhofer Institute for Physical Measurement Techniques IPM, Heidenhofstra\ss e 8, D-79110 Freiburg, Germany}

\author{Ingo Breunig}
\affiliation{Laboratory for Optical Systems, Department of Microsystems Engineering - IMTEK, University of Freiburg, Georges-K\"{o}hler-Allee 102, D-79110 Freiburg, Germany}
\affiliation{Fraunhofer Institute for Physical Measurement Techniques IPM, Heidenhofstra\ss e 8, D-79110 Freiburg, Germany}

\author{Tobias Herr}\email{Corresponding author: tobias.herr@csem.ch}
\affiliation{Swiss Center for Electronics and Microtechnology (CSEM), Time and Frequency, Rue de l'Observatoire 58, CH-2000 Neuch\^{a}tel, Switzerland}

\begin{abstract}
	\normalsize{\textbf{Optical frequency combs are key to optical precision measurements. While most frequency combs operate in the near-infrared regime, many applications require combs at mid-infrared, visible or even ultra-violet wavelengths. Frequency combs can be transferred to other wavelengths via nonlinear optical processes, however, this becomes exceedingly challenging for high-repetition rate frequency combs. Here, it is demonstrated that a synchronously driven high-$Q$ microresonator with a second-order optical nonlinearity can efficiently convert high-repetition rate near-infrared frequency combs to visible, ultra-violet and mid-infrared wavelengths providing new opportunities for microresonator and electro-optic combs in applications including molecular sensing, astronomy, and quantum optics.}}
\end{abstract}

\maketitle %% required

%\setstretch{2} % für doppelten Zeilenabstand

%\section{Introduction}
Optical frequency combs periodically emit ultra-short laser pulses producing an optical spectrum of equally spaced frequencies \cite{Cundiff2003}. Such combs are the backbones of optical precision measurements \cite{Newbury2011}, could boost the bandwidth of optical telecommunication networks and are also vital parts in future optical clocks that are expected to supersede the current definition of the SI second \cite{Diddams2001,Hinkley2013}.
Owing to these prospects, optical frequency comb technology has experienced rapid and significant advances not only in terms of linewidth, power, etc. but also in terms of reduced complexity and size.
For technical reasons such as chromatic material dispersion or availability of gain media and saturable absorbers, frequency combs operate typically in the near-infrared (NIR) spectral regime around 1550~nm wavelength. Many application such as astronomical spectroscopy, quantum physics, optical clocks or molecular sensing however require the frequency comb to be centered in the visible (VIS), ultra-violet (UV) or mid-infrared (MIR) \cite{Diddams2001,Hinkley2013, McCracken2017, Mizrahi2014, Schliesser2012}.  
In order to access these spectral regimes optical frequency conversion via nonlinear parametric processes is employed, transferring the original comb to the wavelength region of interest. Such nonlinear parametric frequency conversion implies energy and momentum conservation for the involved photons ensuring that the converted spectra are still frequency combs with the same comb line spacing as the original comb. Notably, cavity enhanced nonlinear frequency conversion e.g. sub-harmonic generation in synchronously pumped optical parametric oscillators or high-harmonic generation has enabled frequency comb transfer from the NIR to the MIR, VIS and UV spectral regimes \cite{Vodopyanov2011,Muraviev2018,Gohle2005}. 
For efficient nonlinear conversion, pulse energies larger than 1~nJ are common. While feasible in specialized optical research laboratory, such high pulse energies are not readily achievable outside a research context and indeed prohibitive in frequency combs with high-pulse repetition rates in excess of 10 GHz. Here, due to the otherwise excessive average optical power level, pulse energies are typically limited to the pJ-level. Among such high-repetition rate combs are continuous-wave (CW) pumped Kerr-combs and soliton-combs in $\chi^{(3)}$ Kerr-nonlinear microresonators \cite{DelHaye2007, Herr2014}, which are of particular relevance to low-power, compact or mobile applications including time keeping, communication, and optical spectroscopy \cite{Papp2014, Kippenberg2011, Kippenberg2018}. 
Besides efforts towards direct generation at VIS and MIR wavelength \cite{Lee2017,Yu2016}, second and third harmonic generation allowed transferring several Kerr-comb lines into the VIS regime \cite{Jung2014, Wang2016,Fujii2017}. 
Recent work in CW-pumped $\chi^{(2)}$-nonlinear microresonators showed combs in the NIR and their second harmonics \cite{Ikuta2018, Guo2018}. Moreover, it was demonstrated that microresonators permit frequency transfer of a single CW laser frequency to the MIR, UV and VIS, via $\chi^{(2)}$- or $\chi^{(3)}$-nonlinear processes \cite{Ilchenko04,Furst10,Moore11,Breunig16,Meisenheimer17,Carmon2007,Levy2011}. Generally, stringent requirements on chromatic dispersion and phase-matching severely limit the bandwidth of the parametric nonlinear processes.

Here, we explore for the first time synchronously pumped high-$Q$ microresonators with a $\chi^{(2)}$-nonlinearity and their potential for broadband high-repetition rate frequency comb up- and down-conversion. Remarkably, in resonators with engineered phase-matching, we observe that harmonic up-conversion can transfer a broadband NIR comb into the VIS and UV wavelength regime. Moreover, down-conversion via degenerate optical parametric oscillation generates the sub-harmonic of the pump comb; non-degenerate optical parametric oscillation enables wavelength tunable signal and idler combs. 

\begin{figure}[htbp]
	\centering
	\includegraphics[width=8.7cm]{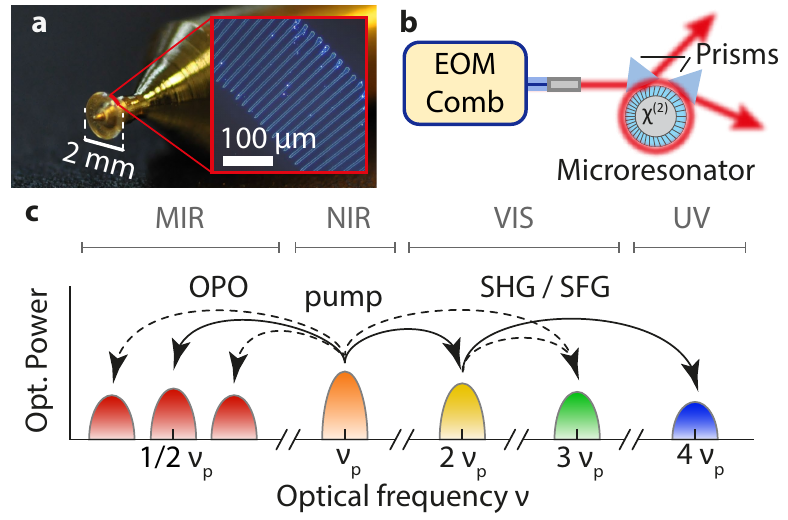}
	\caption{Experimental setup: (a) High-$Q$ $\chi^{(2)}$-nonlinear microresonator on a brass mount. Inset: Dark-field microscope image of a typical quasi-phase matching structure after domain-selective etching. Domain walls appear as bright lines. (b) Electro-optic modulation (EOM) comb coupled to the resonator. Prisms are used for input and output coupling. (c) Nonlinear processes for frequency comb transfer. OPO: optical parametric oscillation, SHG: second harmonic generation, SFG: sum frequency generation, $\nu_\textrm{p}$ denotes the optical center frequency of the pump comb.}
	\label{fig:1}
\end{figure}

%\section{Results}
The high-$Q$ microresonators (Figure~\ref{fig:1}a) are manufactured from $z$-cut 5 \%-MgO-doped congruent lithium niobate wafers. Essential to nonlinear frequency conversion in these resonators is fulfilling the phase-matching condition $m_{\textrm{3}} = m_{\textrm{1}} + m_{\textrm{2}} + \Delta m$, where $m_{i}$ represents the longitudinal mode number of the $i$-th interacting wave and $\Delta m$ indicates the phase mismatch. 
The latter needs to be zero or compensated by a quasi-phase matching (QPM) structure in the resonator. Here, prior to shaping the resonators, such a QPM structure is induced in the wafer material via current-controlled calligraphic electric field poling \cite{Herr18}. The resulting modulation of the $\chi^{(2)}$-nonlinearity along the circumference of the resonator phase-matches processes with non-zero $\Delta m$. 
After poling, the resonators are shaped by means of a femtosecond laser and polished, resulting in $Q$-factors exceeding $10^7$ (critically coupled linewidth of $\approx$~20~MHz at 1550~nm) \cite{Herr18}. Geometrically characterized by major and minor radii of $R$~=~1~mm and $r$~=~0.25~mm, respectively, the resonator has a free spectral range (FSR) of approximately 21~GHz and supports modes with ordinary (o) and extraordinary (e) polarization.

In a first experiment, we attempt up-conversion of a NIR frequency comb. The QPM structure consists of 283 radially oriented domain lines, designed to compensate the phase mismatch $\Delta m = -283$ for second harmonic generation (SHG). Slightly offsetting the center of the QPM structure from the geometric center of the resonator by $\approx$ 100~$\mu$m results in a chirped poling structure enabling both, broadband phase-matching as well as phase-matching of cascaded up-conversion (this is detailed further below).
%\cite{Beckmann11}
Close-to-critical coupling of the fundamental comb to the resonator is achieved by a diamond prism. A second weakly coupled rutile prism permits probing the intracavity field (Fig.~\ref{fig:1}b). For demonstration of frequency comb transfer, the resonator is synchronously pumped by an electro-optic modulation (EOM) frequency comb source \cite{Obrzud2017}, emitting pulses of duration shorter than 0.5~ps at a center wavelength of 1565~nm (cf. Fig.~\ref{fig:1}c). 
%%%fig
The EOM comb allows tuning both, the pulse repetition rate, as well as the comb's offset frequency, so that the pump comb can be made resonant. 
We emphasize that the electro-optic modulation comb is chosen for experimental convenience but can in principle be replaced by any other high-repetition rate comb source including microresonators.
Note that for reduced technical complexity we use sub-harmonic synchronous pumping of the 21~GHz resonators with a 10.5~GHz pulse train \cite{Obrzud2018}. When the initially blue detuned pump lines are tuned into resonance (by changing the pump's offset frequency), the intracavity field builds up and nonlinear optical frequency conversion sets in. Once tuned into resonance, the system is intrinsically stable owing to thermal self-locking without the need for active feedback \cite{Carmon2004}. 

\begin{figure}[htbp]
	\centering
	\includegraphics[width=8.7cm]{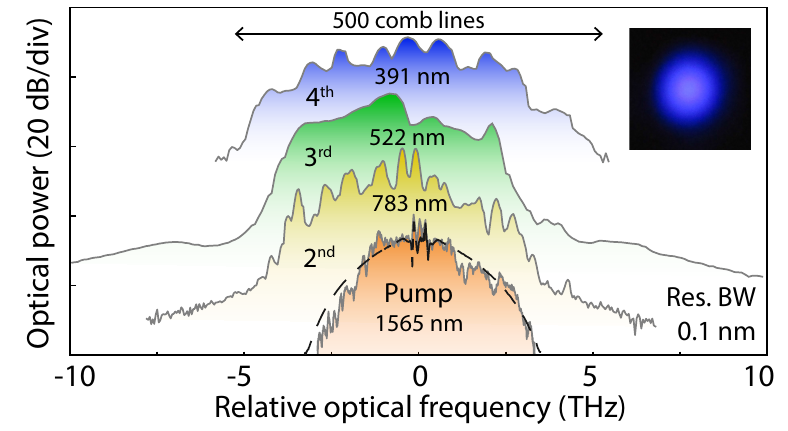}
	\caption{Broadband coupling and comb up-conversion. The dashed black line indicates the spectral envelope of the pump comb before it is coupled to the resonator. Measured optical spectra of the intracavity field at the pump wavelength as well as $2^{\textrm{nd}}$, $3^{\textrm{rd}}$, and $4^{\textrm{th}}$ harmonic. Due to the limited resolution bandwidth of 0.1~nm the individual comb lines are not resolved. Spectra are vertically offset for clarity. Inset: Fluorescence induced by the $4^{\textrm{th}}$ harmonic on a sheet of paper.}
	\label{fig:2}
\end{figure}
Remarkably, when using an average coupled pump power of merely 80~-~100~mW ($\approx$~5~pJ pulse energy) the resonator does not only generate a broadband $2^{\textrm{nd}}$ harmonic spectrum, but indeed broadband $3^{\textrm{rd}}$ as well as broadband $4^{\textrm{th}}$ harmonics of the pump comb as shown in Fig.~\ref{fig:2} comprising up to 500 comb lines in the $4^{\textrm{th}}$ harmonic (recorded by a grating-based optical spectrum analyzer).
The respective out-coupled power in the harmonic combs are 4~mW, 50~$\mu$W and 40~$\mu$W. 
In contrast to the (to the human eye) barely visible $2^{\textrm{nd}}$ harmonic, the $3^{\textrm{rd}}$ harmonic is perceived as a bright, green light emission. 
The invisible ultra-violet $4^{\textrm{th}}$ harmonic, results in strong blue fluorescence when a sheet of paper is brought into the beam (inset in Fig.~\ref{fig:2}).
As illustrated in Fig. \ref{fig:1}c, we suppose that the $2^{\textrm{nd}}$ and $4^{\textrm{th}}$ harmonics result from (cascaded) frequency doubling via sum frequency and second harmonic generation (SFG/SHG), while the $3^{\textrm{rd}}$ harmonic results from SFG of the original comb and the $2^{\textrm{nd}}$ harmonic.
We have no evidence of direct $\chi^{(3)}$-nonlinear third harmonic generation.
%As the $\chi^{(3)}$-nonlinearity in lithium niobate is orders of magnitude smaller than the $\chi^{(2)}$-nonlinearity, and we have seen no evidence of four-wave mixing, we assume that direct $\chi^{(3)}$-nonlinear third harmonic generation is negligible. 
Analysis of the involved polarization shows that the pump comb is o-polarized, the $2^{\textrm{nd}}$, $3^{\textrm{rd}}$, and $4^{\textrm{th}}$ harmonics are e-, o-, and e-polarized, respectively. 
Moreover, while the original comb, $2^{\textrm{nd}}$, and $4^{\textrm{th}}$ harmonic show a TEM00-like mode profile of a fundamental transversal resonator mode, the $3^{\textrm{rd}}$ harmonic is also emitted in higher order modes indicating the involvement of higher-order transverse resonator modes.

\begin{figure}[htbp]
	\centering
	\includegraphics[width=8.7cm]{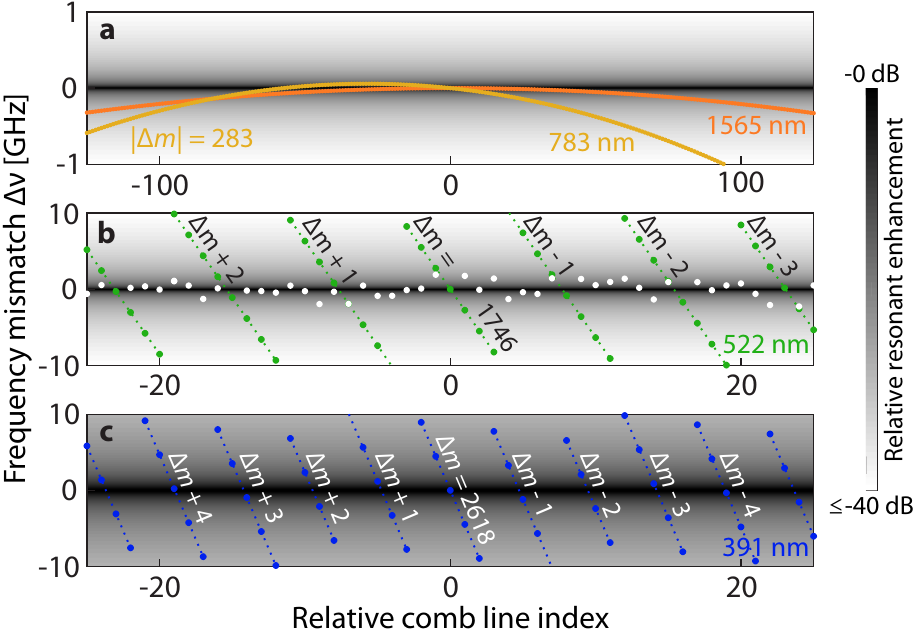}
	\caption{Calculated frequency mismatch $\Delta \nu$ between resonances and comb lines for (a) $1^{\textrm{st}}$, $2^{\textrm{nd}}$, (b) $3^{\textrm{rd}}$, and (c) $4^{\textrm{th}}$ harmonic. The grey shading indicates the relative resonant enhancement in a critically coupled cavity. Full markers represent fundamental transverse modes ($q=1$), white markers in (b) represent higher order transverse modes ($q_{\textrm{3rd}}>1$).}
	\label{fig:3}
\end{figure}

In the following, we investigate how the up-conversion processes can proceed despite chromatic dispersion that implies spectrally varying FSR for a given transverse mode-family. 
To begin with, Figure~\ref{fig:3} relates the resonance width of the critically coupled resonator to the calculated \cite{Gorodetsky2006} frequency mismatch $\Delta \nu$ between resonances and comb lines.
From the graph showing the $\Delta \nu$ for the o-polarized fundamental comb modes, as well as the e-polarized $2^{\textrm{nd}}$ harmonic (Fig.~\ref{fig:3}a), it can be seen that broadband input coupling as well as $2^{\textrm{nd}}$ harmonic generation in o- and e-polarized fundamental transverse modes is possible and compatible with the resonator’s linewidth.
We note that due to the strong depletion of the pump comb as well as the $2^{\textrm{nd}}$ harmonic due to further up-conversion, the effective linewidth of the resonator will be larger than the critically coupled low-power linewidth of 20~MHz. Next, Figure~\ref{fig:3}b shows that $\Delta \nu$ for the $3^{\textrm{rd}}$ harmonic is significantly larger. Here, the green filled dots represent the fundamental transverse modes (with radial mode number $q_{\textrm{3rd}}=1$), which are only sparsely resonant. Moreover, the QPM structure needs to compensate for $\Delta m=1746$, which in addition changes by 1 every 7-8 comb lines. If only the fundamental transverse mode were populated this would result in a sparse comb. However, considering higher transverse optical modes (up to $q_{\textrm{3rd}}=10$) there is always a mode that is close to resonance, as illustrated by the white circles in Figure~\ref{fig:3}c, explaining the multi-modal emission profile of the $3^{\textrm{rd}}$ harmonic light. 
Finally, regarding the $4^{\textrm{th}}$ harmonic ($\Delta m$ values around 2618), the resonance width is significantly increased due to the higher absorption in the UV regime (critically coupled linewidth was measured to be $\approx$~0.8~GHz using a frequency-doubled CW titanium-sapphire laser). Therefore the resonance condition is much less stringent and the comb can effectively be populated in the fundamental transverse mode with $q_{\textrm{4th}}=1$, resulting in a single-mode TEM00-like emission profile (Fig.~\ref{fig:2}).

\begin{figure}[htbp]
	\centering
	\includegraphics[width=8.7cm]{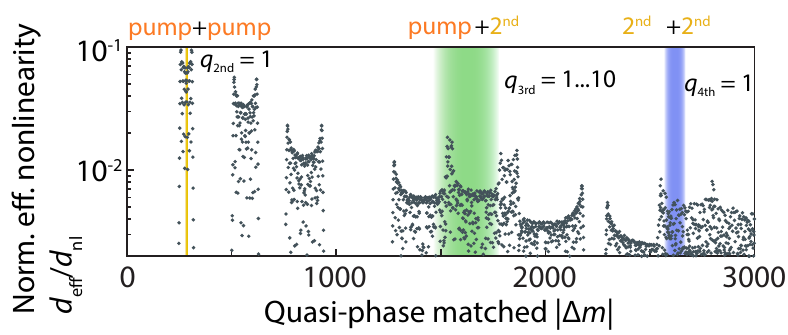}
	\caption{Analysis of quasi-phase matching structure. The normalized effective nonlinearity is shown as a function of $\Delta m$ (grey markers). Color-shaded areas indicate the required $\Delta m$ values for the respective nonlinear conversion processes for 500 comb lines with $q_i$ being the radial mode number of the $i$-th harmonic.}
	\label{fig:QPM}
\end{figure}

As illustrated in Figure~\ref{fig:3}, the QPM structure is required to compensate for a large number of different $\Delta m$. 
%In order to understand how the QPM structure can support harmonic generation, the QPM structure is analyzed assuming a realistic domain-width-to-period-length ratio of 25 \%. 
The range of compensated $\Delta m$ values can be found via Fourier-analysis of the QPM structure \cite{Hum2007}.
Figure~\ref{fig:QPM} shows that the applied QPM structure, assuming a realistic domain-width-to-period-length ratio of 25~\%, is indeed well suited to compensate for all $\Delta m$ values relevant for the generation of $2^{\textrm{nd}}$, $3^{\textrm{rd}}$, and $4^{\textrm{th}}$ harmonic owing to its off-centered design.

\begin{figure}[htbp]
	\centering
	\includegraphics[width=8.7cm]{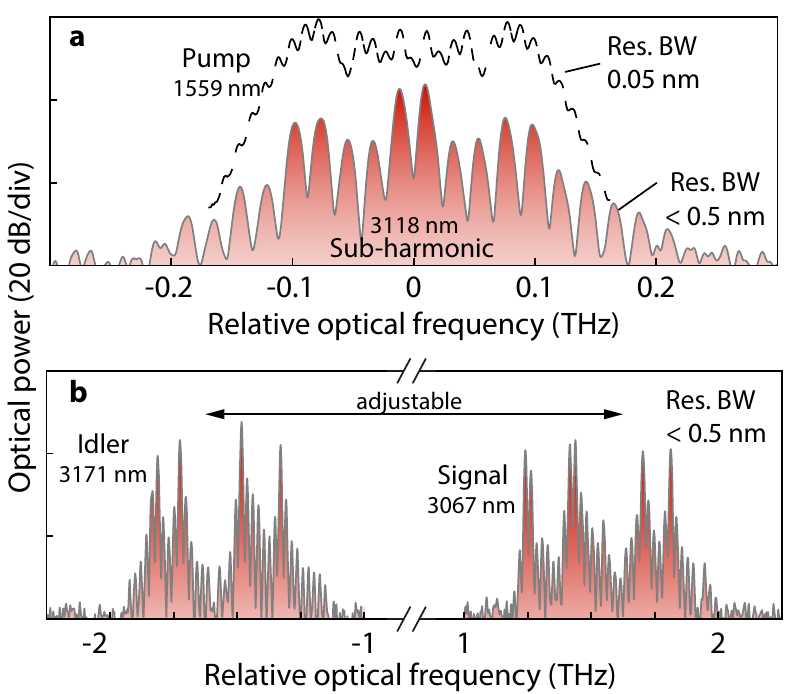}
	\caption{Frequency comb down-conversion: (a) Degenerate, sub-harmonic MIR frequency comb. (b) Non-degenerate MIR comb generation. Signal and Idler separation is adjustable (cf. main text). Comb lines are resolved.}
	\label{fig:4}
\end{figure}

In a second experiment, we explore frequency comb down-conversion into the MIR wavelength regime. 
To this end a similar resonator with a linear QPM pattern with a period of 30.5~$\mu$m is used, again resulting in a chirped structure and broadband phase-matching. Surprisingly, when pumping the resonator with the spectrum shown in Fig.~\ref{fig:2} (center wavelength 1559~nm), several competing optical parametric down-conversion processes with corresponding signal and idler combs are observed. As careful analysis of the optical spectra reveals, this large number of signal and idler combs does not represent a single frequency comb but a number of combs with generally different offset frequencies. 
This is similar to sub-comb formation in $\chi^{(3)}$-nonlinear microresonators \cite{Herr2012}. In order to restrict the occurring nonlinear processes to only one pair of signal and idler combs we reduce the spectral bandwidth of the driving comb to approximately 0.4~THz. 
For a suitable pump to resonance detuning the generated signal and idler combs are degenerate and form a single consistent sub-harmonic frequency comb at exactly half the optical frequency of the original comb (cf. Fig.~\ref{fig:4}a).
Changing the pump comb's offset detuning, its center wavelength, or the temperature of the microresonator leads to the generation of non-degenerate signal and idler combs whose frequency separation can be modified (Figure \ref{fig:4}b). We anticipate that future experiments in on-chip $\chi^{(2)}$-nonlinear resonators with reduced number of transverse modes \cite{Jung2014,Wolf2018} will significantly reduce the number of competing optical parametric processes, permitting the generation of broader spectra.
Finally, we note that while two different resonators have been used for up- and down-conversion, this could be implemented in a single microresonator, phase coherently transferring and linking frequency combs across MIR, NIR, VIS and UV wavelength domains. 

%\section{Conclusion}
In conclusion, we have demonstrated the first synchronously pumped $\chi^{(2)}$-nonlinear microresonator, enabling frequency comb transfer from the NIR to MIR, VIS, and UV wavelengths. This provides novel opportunities for high-repetition rate frequency combs (e.g. microresonator-based combs and EOM combs) in spectroscopy, sensing, navigation, and time keeping. Besides providing access to challenging wavelength regimes, the possibility of transferring frequency combs to MIR and UV frequencies in the same microphotonic device platform could enable low-power self-referenced microphotonic frequency combs without the need for octave spanning spectra (e.g. via 3f-4f interferometry).

\section*{Funding}
This work was supported by the Swiss National Science Foundation (20B2-1\_176563), the Canton of Neuch\^{a}tel, the Fraunhofer and Max Planck cooperation programme, and the Bundesministerium f\"{u}r Bildung und Forschung (13N13648). YJ acknowledges support by the Alexander von Humboldt Foundation.

%\section*{Supplemental Documents}

%\bigskip \noindent See \href{link}{Supplement 1} for supporting content.

%\section*{References}

% Bibliography
\bibliography{comb_citations}

% Note that this extra page will not count against page length.
%\bibliographyfullrefs{comb_citations}

\end{document}